# Sorting Network for Reversible Logic Synthesis

*Md. Saiful Islam, Md. Rafiqul Islam[\*], Abdullah Al Mahmud and Muhammad Rezaul karim*

*Department of Computer Science & Engineering*
*University of Dhaka, Dhaka-1000, Bangladesh*

[\*]*School of Information Technology*
*Deakin University, Melbourne, Australia*

{sohel_csdu, rafik3203, aamrubel, r_karimcs}@yahoo.com

## Abstract

*In this paper, we have introduced an algorithm to implement a sorting network for reversible logic synthesis based on swapping bit strings. The algorithm first constructs a network in terms of n*n Toffoli gates read from left to right. The number of gates in the circuit produced by our algorithm is then reduced by template matching and removing useless gates from the network. We have also compared the efficiency of the proposed method with the existing ones.*

## 1. Introduction

A reversible circuit maps each input vector into a unique output vector. Landaur's principle [1] proved that logic computations that are not reversible necessarily dissipate heat irrespective of their implementation technologies. It is shown that zero energy dissipation would be possible only if the network consists of reversible gates [2]. Thus reversibility will become an essential property in future circuit design. Synthesis of reversible logic circuits differs significantly from the synthesis of classical logic circuits. Because in a reversible circuit the number of inputs must be equal to the number of outputs, every output can be used only once and the resulting circuit must be acyclic.

An *n*-input *n*-output totally specified Boolean function $f(X)$, $X = \{x_1, x_2, ..., x_n\}$ is reversible iff it maps each input assignment to a unique output assignment.

A reversible function can be written as a standard truth table as in Table 1 and can also be viewed as a bijective mapping of the set of integers $0, 1, ..., 2^n - 1$. Hence a reversible function can be defined as an ordered set of integers corresponding to the right side of the table, e.g. {1,0,3,2,5,7,4,6} for the function in Table 1. We can thus interpret the function over the integers as $f(0) = 1$, $f(1) = 0$, $f(2) = 3$, etc.

An *n*-input *n*-output gate is reversible if it realizes a reversible function. Many reversible gates have been proposed in the literature. One of the first gates was the CNOT gate [3], which capable of producing the "exclusive or" of two input bits as the second output and the first output is equal to the first input.

| c b a | c° b° a° |
|-------|----------|
| 0 0 0 | 0 0 1 |
| 0 0 1 | 0 0 0 |
| 0 1 0 | 0 1 1 |
| 0 1 1 | 0 1 0 |
| 1 0 0 | 1 0 1 |
| 1 0 1 | 1 1 1 |
| 1 1 0 | 1 0 0 |
| 1 1 1 | 1 1 0 |

Table 1. 3*3 Reversible logic function

A generalization of CNOT is a 3-input 3-output Toffoli gate [4]. The Toffoli gate negates the third bit *iff* the first two bits are 1. Figure 1 shows both gates as they are commonly drawn.

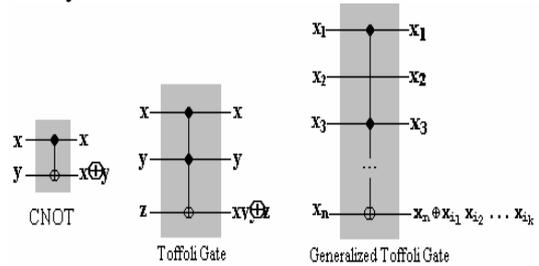

Figure 1. CNOT and Toffoli gate

Figure 2. Generalized Toffoli gate

A generalized *n*n* Toffoli gate changes one bit, called the target, if some of the *k* bits are 1 which is shown in Figure 2. The changing bit, also called target, may also be in any position. The gate will be defined as follows $T(x_{i1}, x_{i2}, ..., x_{ik}: x_n)$ where $x_n$ is the target and $x_{i1}, x_{i2}, ..., x_{ik}$ are the control bits.

Garbage is the number of outputs added to make an *n*-input *k*-output function ($(n, k)$ function) reversible.

Given two bit strings, *P* and *Q*, the Hamming distance between them, denoted $\delta(P, Q)$ is the number of positions for which *P* and *Q* differ.

***Example 1.*** Consider the bit strings (1,0,1) and (0,1,1). The Hamming distance between these two bit strings is 2 since the number of positions for which these two bit strings differ is 2.

Given the function $f(X)$, the complexity $C(f)$ is defined as the sum of the individual Hamming distances over the $2^n$ input-output patterns. For example, the value of $C(f)$ for the function in Table 1 is 8.

*Lemma 1.* In any reversible specification the upper and lower bound on the Hamming distance, δ, between any two bit strings P and Q, is $n$ and 1, where $n$ is the number of input lines. That is,
$$1 \leq \delta(P, Q) \leq n.$$

*Proof:* Let $P$ be $(a_1, a_2,...,a_n)$ and $Q$ be $(b_1,b_2,...,b_n)$. Since in a reversible specification no two bit strings are identical, they must differ in at least one position. Let $m$ be the index at which $a_m \neq b_m$. That is, $a_m = b_m'$, and $b_m$ is either 0 or 1. Bit strings $P$ and $Q$ may differ at most every position, i.e., $a_i \neq b_i$, where $1 \leq \delta \leq n$. Thus we can conclude that $1 \leq \delta(P, Q) \leq n$. €

*Lemma 2.* Two bit strings $P$ and $Q$ can be swapped without affecting others *iff* the Hamming distance between them is 1.

*Proof:* In any reversible specification bit string P and Q will be unique. Let P be $(p_1, p_2, ..., p_n)$ and Q be $(q_1, q_2, ..., q_n)$. Also let m be the index for which $p_m \neq q_m$. That is, $p_i = q_i$, where $1 \leq i \leq n$ and $i \neq m$. Since no bit string except P & Q will contain $(p_1, p_2, ..., p_{m-1}, x, p_{m+1}, ... p_n)$ where x may be either '0' or '1'. If we use $(p_1, p_2, ..., p_{m-1}, p_{m+1},..., p_n)$ as the control bits to drive the Toffoli gate, only P & Q will be affected. €

*Example 2.* In Figure 3, bit strings (1,1,1) and (1,1,0) have been swapped using T(b,c:a). Since the Hamming distance between them is one, this swapping can be carried out without affecting others.

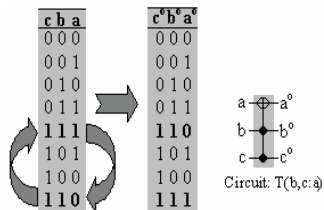

Figure 3. Swapping bit strings

## 2. Sorting network

Section 1 has described that a reversible function can be defined as an ordered set of integers corresponding to the right side of the table, e.g. {1,0,3,2,5,7,4,6} for the function in Table 1. Therefore, if we can build a network of reversible gates that might sort this set, it will eventually realize the function. For example, the ordered set of integers for the function in table 1 will become {0,1,2,3,4,5,6,7} and the index of each element will be equal to itself. That is, if we define the set as $\{p_i\}$, then $2^n-1 \geq i \geq 0$ and $p_i = i$ where n is the number of input lines. Section 2.1, 2.2, 2.3 and 2.4 will present such an algorithm named **BSSSN** (**Bit String Swapping Sorting Network**) that will construct a network as a sequence of Toffoli gates that might sort the elements of the set. The idea here is based on swapping bit strings whose hamming distance is exactly one as described in section 1.

### 2.1. Output translation

**BSSSN:** *Constructing the sorting network and its corresponding circuit*

While (there is a bit string $(a_1,a_2,...,a_n)$ in the set that is not in its intended place, i.e., int_value$(a_1,a_2,...,a_n) \neq$ index)

1. Let $(a_1,a_2,...,a_n)$ is not in its intended place
   then by induction its place is occupied by
   another string, say $(b_1,b_2,..., b_n)$
2. Compute dist = δ $((a_1,a_2,...,a_n), (b_1,b_2,...,b_n))$
   if dist=1, swap $(a_1,a_2,...,a_n)$ and $(b_1,b_2,...,b_n)$
      using gate T$(a_1,a_2,..., a_{k-1},a_{k+1}, ...,a_n : a_k)$
         where $a_k \neq b_k$
   else{
      find all the bit strings $\{(c_1,c_2, ... ,c_n)\}$
      such that δ $((b_1,b_2,...,b_n) , (c_1,c_2, ... ,c_n))$=1
      swap $(b_1,b_2,...,b_n)$ with one of the $(c_1,c_2,... ,c_n)$
      such that δ $((a_1,a_2,...,a_n), (c_1,c_2, ... ,c_n))$
      will be minimum and in case of multiple
      $(c_1,c_2, ... ,c_n)$ low int_value bit string is
      chosen to break the tie and that is not in its
      intended place }
3. Now $(c_1,c_2, ... ,c_n)$ will become the new
   $(b_1,b_2,...,b_n)$ and goto step 2.

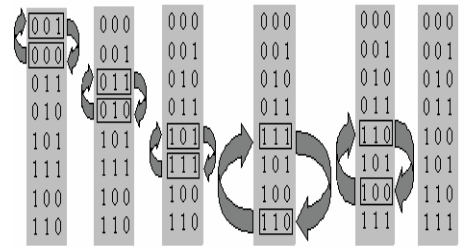

Figure 4. Constructing network using BSSSN

Figures 4 and 5 illustrate the application of BSSSN. The sequence of gates that form the network is: T(b2,c2:a) T(b,c2:a)T(a,c:b)T(b,c:a)T(a2,c:b).

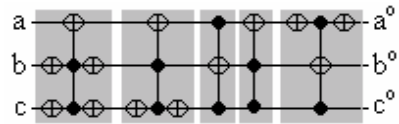

Figure 5. Circuit for the network constructed by BSSSN

Algorithm BSSSN is straightforward. It is greedy in the sense it hopes that a bit string can be swapped by the one that is in its intended place. Because of Lemma 1 and 2, it is always possible to find two bit strings for Steps 2 that can be swapped. Therefore, it will always terminate successfully with a circuit for a given specification. The best case occurs when a bit string can always be swapped by the bit string placed in its intended position. A variant of BSSSN takes the bit string in the set whose integer representation is low and brings it to its intended place. Figures 6 and 7 illustrate the application of the variant of BSSSN and the sequence of gates that form the network is: T(b2,c2:a)T(b,c2:a)T(b2,c:a)T(a,c:b) T(b,c:a).

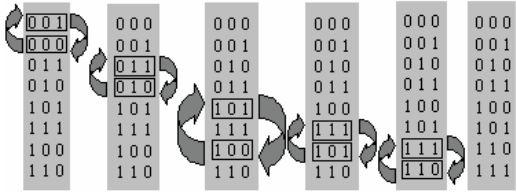

Figure 6. Constructing network using variant of BSSSN

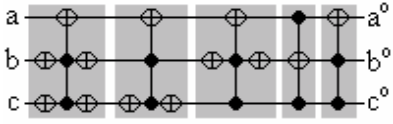

Figure 7. Circuit for the network constructed by variant of BSSSN

## 2.2. Input translation

For input translation, we have to find an inverse of the specification. For example, the reverse specification of the function in Table 1 is {1,0,3,2,6,4,7,5}. Then, we can apply BSSSN to realize the function.

By applying BSSSN to realize the reverse specification we get the circuit: T(b2,c2:a)T(b,c2:a)T(b,c:a)T(a,c:b)T(b2, c:a) and its variant, we get: T(b2,c2:a)T(b, c2:a)T(a2,c:b) T(b,c:a)T(a,c:b)T(b,c:a).

## 2.3. Random selection and control input reduction

To sort the elements in a specification, algorithms BSSSN and its variant take one element at a time and bring it to its intended place. Selection of a bit string randomly and then returning to its intended place can reduce the total number of swapping. This will minimize the total number of gates in the network also.

Algorithm BSSSN also assigns the maximum number of control lines to each Toffoli gate. For larger problems with up to 8 or 9 inputs this may not be a practical one. Selective use of control inputs can be used to swap elements. This can be carried out safely as long as it will not affect bit strings that are already in its intended place. We should choose a subset of the control inputs that will minimize the $C(f)$ of the resulting specification. For example, we can select control inputs that will drive a Toffoli gate to bring more than one element at a time to their intended place.

## 2.4. Reduction rules

The circuits produced by BSSSN as described thus far frequently have gate sequences that can be reduced. For example, the sequence T(b:a) T(:b) T(:a) can be replaced by the sequence $T(:b)$ T(b:a). Here we have implemented template driven reduction method introduced in [5].

In addition to template matching in [5] we have also removed useless gates that come in pairs and have no effect in the circuit. For example, T(a,b:c) is an example of a useless gate in the gate sequence T(a,b:c) T(a:c)T(a,b:c) and the gate sequence can be replaced by T(a:c) without any modification in the circuit. When identifying useless gate in a circuit, we have taken into account of Property 2.2 which follows directly from the definition of $n*n$ Toffoli gates.

**PROPERTY 2.2:** A gate $T(x_1,x_2,...,x_{k-1}: x_k)$ can be removed from the sequence $T(x_1,x_2,...,x_{k-1}: x_k)$ $T(a_1,a_2,...,a_{l-1}: a_l)T(b_1,b_2,...,b_{m-1}:b_m)$ …$T(c_1,c_2 ,... ,c_{n-1}: c_n)$ $T(x_1,x_2,...,x_{k-1}: x_k)$ iff $x_k$ ∉ {$a_1,a_2,..., a_{l-1,} b_1,b_2,...,b_{m-1},...,c_1,c_2 ,... ,c_{n-1}$} and $a_l,b_m,...,c_n$ ∉ {$x_1,x_2,...,x_{k-1}$}.

## 3. Experimental results

In Section 3.1 we have shown some reversible examples and compare them with the circuits in [6][7]. Section 3.2 describes the method used to convert an irreversible specification to a reversible one and the way to synthesize them. Here we have synthesized a reversible circuit from an irreversible specification and not transformed an irreversible circuit to a reversible one.

## 3.1. Reversible examples

For each example, the specification is given as an ordered set of integers, which define the truth table specification of the reversible logic function to be realized. The circuit is given as an ordered sequence of Toffoli gates. Read from left to right they transform the left side to the right side.

**Example 3.1** Verification of realizing a **Fredkin gate.** This example is collected from [7]. The circuit given by our method produces the same result.

*Specification:* {0,1,2,3,4,6,5,7}
*Circuit(BSSSN)* : T(a:b) T(b,c:a)T(a:b)
*Circui(VAR_ BSSSN)* : T(b:a)T(a,c:b)T(b:a)

**Example 3.2** This is a second example of the interchange of two positions in the specification. The circuit given by our method is identical to the solution provided by [8].

*Specification:* {0,1,2,4,3,5,6,7}
*Circuit(BSSSN)*: T(a,b:c)T(a,c:b)T(b2,c:a)T(a,c:b) T(a,b:c).
*Circui(VAR_ BSSSN)*:
 T(a2,b2:c)T(b2,c2:a)T(a,c2:b)T(b2, c2:a) T(a2,b2:c).

**Example 3.3** This example is taken from [6]. The circuit given by our method is identical to the solution provided by the Bidirectional Algorithm in [6].

*Specification:* {7,0,1,2,3,4,5,6}
*Circui(VAR_ BSSSN)*: T(a,b:c)T(a:b)T(:a).

## 3. 2. Non-reversible examples

According to [9] an irreversible function can be realized using reversible gates with the addition of some number of constant inputs and 'garbage' outputs. The minimum number of garbage outputs required to transforming an irreversible function to a reversible one is ⌊$\log_2^m$⌋, where m is the maximum number of times a single output pattern appears in the specification.

A single-output or a multi-output function $f$ involving input variables $x_1$, $x_2$, ...,$x_n$ can be transformed to a reversible specification in the following way.

a. Compute *m*, where *m* is the maximum output pattern multiplicity of the irreversible specification. Let $p = \lceil \log_2^m \rceil$ and *k* is the number of outputs of *f*. The value of *k* will be one for single-output function.
b. If $(p + k) > n$
  i. Add $(p + k - n)$ new input variable $x_{n+1}, x_{n+2}, ..., x_{p+k}$ and set each of them to zero on input in the circuit.
  ii. Add n outputs each equal to one of the original inputs $x_1, x_2, ..., x_n$.
  iii. *k* outputs will be realized on $x_{p+1}, x_{p+2}, ..., x_{p+k}$.
  iv. for single-output function replace *f* by $f \oplus x_{n+1}$.
Else
  i. Add *n-k* outputs each equal to one of the original inputs $x_1, x_2, ..., x_{n-k}$.
  ii. The *k* outputs will be realized on inputs $x_{n-k+1}, x_{n-k+2}, ..., x_n$.

It is easily verified that the specification constructed in this way maps an input pattern to a unique output pattern and is therefore reversible. The approach used by Miller *et al.* [7] adds unnecessary constant inputs and thus produces extra garbage outputs that are not actually needed to make the specification reversible.

**Example 3.4** This procedure for transforming a single-output function is illustrated for the example of the 2-input EX-OR function in Table 2. The resulting circuit is the single gate *T(a:b)* which realizes the EX-OR of *a* and *b* on *b*. But the same function is realized in [7] with one constant on inputs and thus produces one extra garbage output that is unnecessary.

| b a | f | b a | b° a° | c b a | c° b° a° |
|---|---|---|---|---|---|
| 0 0 | 0 | 0 0 | 0 0 | 0 0 0 | 0 0 0 |
| 0 1 | 1 | 0 1 | 1 1 | 0 0 1 | 1 0 1 |
| 1 0 | 1 | 1 0 | 1 0 | 0 1 0 | 1 1 0 |
| 1 1 | 0 | 1 1 | 0 1 | 0 1 1 | 0 1 1 |
|  |  |  |  | 1 0 0 | 1 0 0 |
|  |  |  |  | 1 0 1 | 0 0 1 |
|  |  |  |  | 1 1 0 | 0 1 0 |
|  |  |  |  | 1 1 1 | 1 1 1 |
| (a) |  | (b) |  | (c) |  |

Table 2. (a) 2-input EX-OR (b) reversible specification derived from 2-input EX-OR using method described above (c) using method in [7]

**Example 3.5** This example illustrates the realization of 2-input AND function. The specification is {0,1,2,7,4,5, 6,3}. The solution produced by our algorithm is same to the solution provided by that of the [7], that is, T(a,b:c) which realizes the AND of *a* and *b* when *c* is 0 on input.

**Example 3.6 Full Adder Minimization:** The resulting circuit produced by our algorithm is:
T(a,b2,d2:c)T(a2,b,d2:c) T(a,b,c2:d)
T(a,b2,d:c)T(a,b2,c:d)T(a2,b,d:c) T(a2,b,c:d)T(a,b,c:d).

which can be simplified by template matching to T(:d) T(a,b,d:c) T(a,d:c) T(a,b,d:c) T(b,d:c) T(:d)T(a,b,c:d)T(a,b:d)T(a,b,d:c)T(a,d:c)T(a,b ,c:d)T(a,c:d)T(a,b,d:c)T(b,d:c) T(a,b,c:d)T(b,c:d)T(a,b,c:d).

which can be again simplified by removing useless gates to

T(:d) T(a,d:c) T(b,d:c) T(:d) T(a,b,c:d) T(a,b:d)T(a,b,d:c) T(a,d:c)T(a,b,c:d)T(a,c:d)T(a,b,d:c)T(b,d:c)T(b,c:d).

After final simplification we get the following circuit:

T(a,b:d)T(a:b)T(b,c:d)T(b:c).

This circuit is identical to the circuit in [6][7].

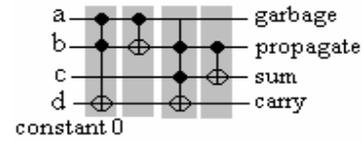

Figure 8. Full Adder

Though the initial circuit produced by our algorithm seems to be larger one, it can be simplified easily by simple template matching and identifying useless gates. Thus the circuit will be optimal. The main advantage of our algorithm is that it does not require exhaustive analysis like spectral used in [7].

## 4. Conclusions

An algorithm to realize totally specified reversible specification has been presented. The algorithm always terminates with a network of Toffoli gates that can translate both input and output side to their corresponding output and input side. Since the synthesis of reversible circuits can be done in either side, this is valid.